# Scaling properties of gelling systems in non-linear shear experiments

**A. Louhichi[1*], M-H Morel[2], L. Ramos[1], A. Banc[1*]**

[1] *Laboratoire Charles Coulomb (L2C), Univ. Montpellier, CNRS, Montpellier, 34095 France.*

[2] *UMR IATE, Université de Montpellier, CIRAD, INRAE, Montpellier SupAgro, 2 pl. Pierre Viala, 34060 Montpellier, France.*

*Email : amelie.banc@umontpellier.fr*

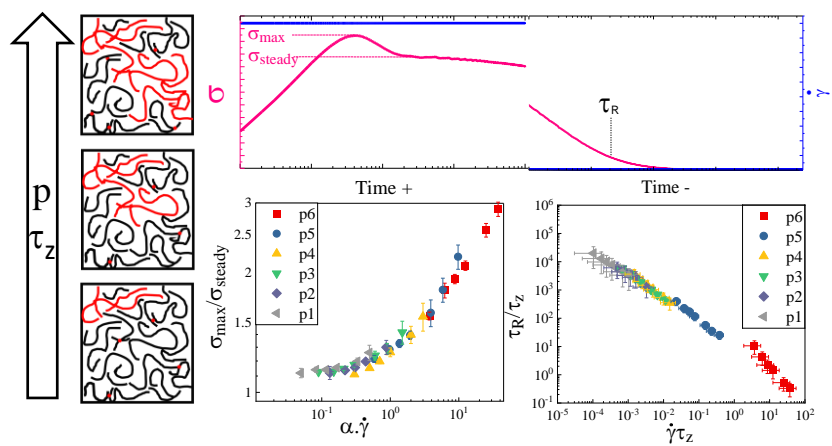

## Abstract:

We study model near-critical polymer gelling systems made of gluten proteins dispersions stabilized at different distances from the gel point. We impose different shear rates and follow the time evolution of the stress. For sufficiently large shear rates, an intermediate stress overshoot is measured before reaching the steady state. We evidence self-similarity of the stress overshoot as a function of the applied shear rate for samples with various distances from the gel point, which is related to the elastic energy stored by the samples, as for dense systems close to the jamming transition. In concordance with the findings for glassy and jammed systems, we also measure that the stress after flow cessation decreases as a power law with time with a characteristic relaxation time that depends on the shear rate previously imposed. These features revealed in non-linear rheology could be the signature of a mesoscopic dynamics, which would depend on the extent of gelation.

Gelation is a liquid-to-solid transition of polymeric or colloidal solutions, which is induced by the progressive formation of either covalent or physical bonds as crosslinks. At the sol-gel transition, samples pass from a pre-gel state to a post-gel state through a critical state that coincides with the apparition of a percolated space-spanning network. At the critical point, gels are characterized in the linear regime by a power law rheology, where identical power law exponents characterize the frequency dependence of the storage, $G'$, and loss, $G''$, moduli: $G' \propto G'' \propto \omega^n$. The exponent $n$ is comprised between 0 and 1 and is associated with the fractal dimension of a hierarchical network structure[1]. In the vicinity of the gel point, the samples display the same power law rheology at high frequency but behave as simple viscous liquid or elastic solid at low frequency depending on their relative position to the gel point. The transition between the two frequency regimes is delineated by the longest relaxation time $\tau_z$, which truncates the power-law decay of the spectrum of relaxation times associated to the power law rheology. Pre-gel samples are characterized by a self-similar distribution of fractal clusters, and as the extent of the crosslinking reaction increases, the maximum relaxation time of the distribution of clusters, $\tau_z$, increases and diverges at the critical state. Beyond this point, the self-similar distribution of relaxation times is associated with the network mesh size distribution and $\tau_z$ decreases as the percolated network densifies. During the gelation process, a superposition of linear viscoelastic functions was evidenced and qualified as a time-cure superposition[2].

To date, the theoretical framework for the linear viscoelasticity of near critical gels is well-established and has been experimentally checked for a large variety of materials such as synthetic polymers[2–4], biopolymers[5,6], polypeptides[7], proteins[8–11], nanocrystals[12] and clays[13,14]. However, data in the non-linear regime are much scarcer[15], and their rationalization is partial. Analysis of start-up shear experiments for near critical gels were often limited to qualitative evaluations of the strain softening/hardening of the samples [16–18]. A modelling attempt was published but it was limited to one gelation state beyond the critical point[10]. Securing a fundamental understanding of the non-linear behavior of near-critical gels is however of crucial importance, because of their everyday use and/or processing that involves large strains and a wide range of strain rates.

Whereas for most gelling samples the critical state is transient and evolves rapidly compared to the experimental time, some systems stabilize close to the gel state, allowing detailed investigation in the non-linear regime. This is the case with some daily used materials such as pressure-sensitive adhesives[19], chewing-gums[16] or gluten[8,11]. Gluten is the water-insoluble network formed in wheat flour dough by the wide distribution of grain storage proteins. These proteins are intrinsically disordered and behave physically as polymers[20]. They interact mainly through intermolecular disulfide bonds and hydrogen bonds. These bonds can be considered transient due to their weak energy (as compared to strong covalent bonds) and their capacity to reshuffle. Here, we use start-up shear and flow cessation protocols to probe the non-linear viscoelastic behavior of near-critical gelling systems before the gel point, deformed at different rates. We employ well-characterized model gluten samples [8,20–24] for which the distance from the critical gelation point can be finely tuned, allowing a detailed investigation of the impact of the extent of gelation on the response of near-critical gelling systems beyond the linear regime. We show that master curves can be successfully built for the maximum stress, the stress overshoot amplitude during start-up shear experiments, and for the characteristic relaxation time after shear cessation, using linear viscoelastic characteristics of the samples ($\tau_z$ and $n$), thus evidencing the preservation of the network self-similarity also in the non-linear regime. Interestingly, we also find similarities with systems close to the glass or the jamming transitions in the non-linear regime, which we discuss.

Gluten proteins dispersed in water at 500g/L form near-critical gels that slowly evolve on a time scale of days[8]. The sample gelation is mainly controlled by the proportion of glutenin, the polymeric fraction of gluten proteins. Gelation is induced by the hydrogen bonding of chains through a mechanism involving glutenin disulfide bonds reshuffling[8,23]. Here, the extent of gelation of samples of a same age (p) is tuned by sample composition (Table 1): we vary the glutenin content and the amount of N-ethyl maleimide (NEM), a thiol blocker that scavenges free thiols and reduces the rate of thiol-disulfide bonds exchange.

**Table 1.** Sample description. The series of samples without NEM but with different glutenin contents appear in bold. Samples are ordered according to their gelation extent as quantified by $\tau_z$, the characteristic relaxation time measured by linear rheology.

| Sample name  Extent of gelation | Glutenin content  w/w (%) | N-ethyl Maleimide/Free thiol molar ratio |
|---|---|---|
| p1 | 52 | 10 |
| **p2** | **48** | **0** |
| p3 | 52 | 1 |
| p4 | 52 | 0.3 |
| **p5** | **52** | **0** |
| **p6** | **60** | **0** |

To probe the extent of gelation of the different samples, we measure their linear viscoelasticity through a dynamic frequency sweep (DFS) and check the theoretical expectations for near-critical gels[2]. A pre-gel master curve can be built with the DFS measured for all samples using $\tau_z$ and $b_L$ as shifting factors for the frequency and the moduli, respectively (Fig. 1a) (see Fig.

SI1 & Fig. SI2 for alternative representations of these data). The parameter $\tau_z$ defines the longest characteristic time of the power law regime and can be attributed to the largest clusters in the sample. On the other hand, $b_L$ is an effective compliance, a factor proportional to $\frac{\tau_z}{\eta_0}$ with $\eta_0$ the sample zero-shear viscosity[2]. The relationship between these factors (Fig. 1b) satisfies the predictions of the near-critical gel theory[2]: $b_L \propto \tau_z^n$ with $n = 0.4$ the critical exponent of the gel that corresponds to the exponent of the power law of $G'$ and $G''$, at high frequency in the DFS master curve (Fig. 1a). Hence, the series of gluten pre-gel samples satisfy the time-curing superposition with characteristic times $\tau_z$ spanning over 5 decades of frequency. In the non-linear regime, the samples are probed by a succession of shear start-up measurements at different rates, from 0.1 to 10 s$^{-1}$. The experiments at different rates are separated by stress relaxation steps of 2400 s each. This duration is found sufficiently long to allow the shear stress to fully relax before starting the next shear start-up experiment. At the end of the protocol (Fig.1c), a DFS measurement is performed to check the sample's integrity. As shown in Fig. 1a,b, the near-critical gel state of samples is maintained despite a weak ageing during the duration of a full series of measurements (17 hours). This finding suggests that the samples have the ability to fully relax the stress and erase all traces of the deformation history, which aligns with earlier studies on the weak damping of near critical gels[17].

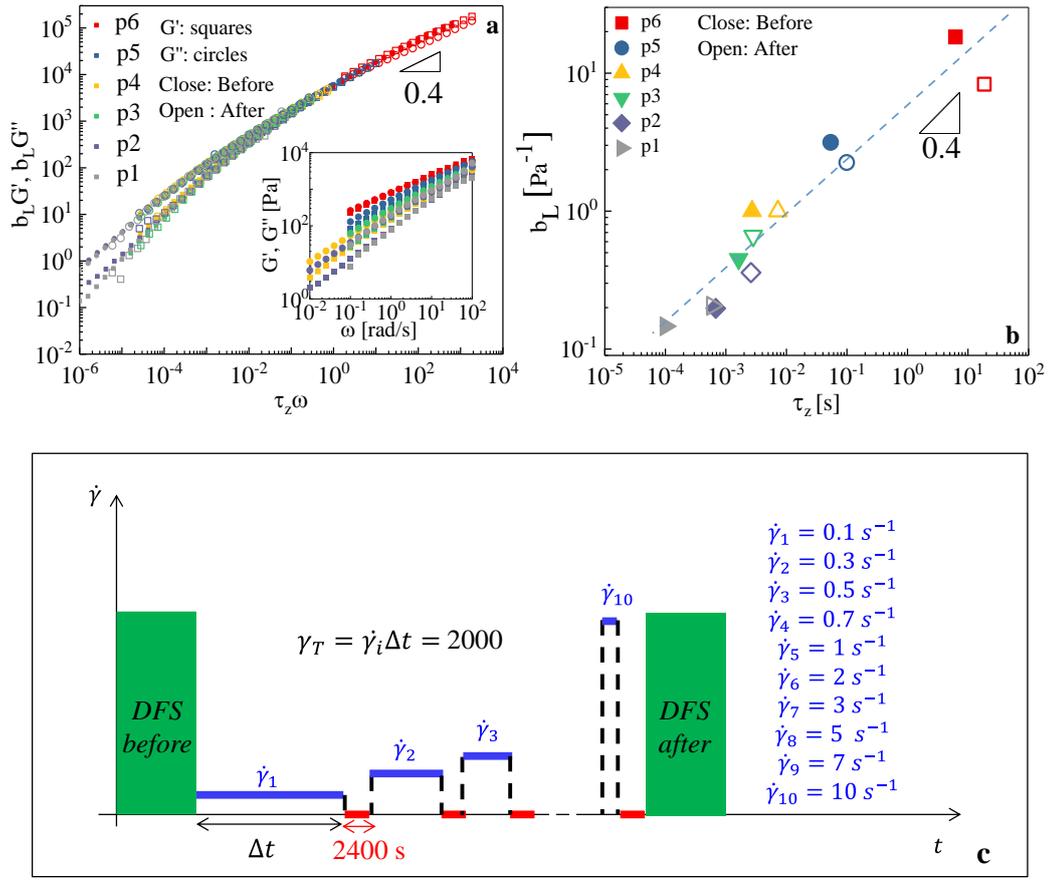

**Figure 1.** (a) Master curve for the storage, G', and loss, G'', moduli as a function of frequency, using horizontal ($\tau_z$) and vertical shifting factors ($b_L$), gathering all samples, as indicated in the legend. Full and empty symbols correspond to data measured respectively before and after start-up shear. Inset: unshifted data measured before the shear start-up protocol. (b) Plot of $b_L$ as a function of $\tau_z$ before and after the whole shear start-up protocol. Symbols are experimental data points, and the dashed line is a power law fit yielding an exponent 0.4. (c) Experimental protocol applied to each sample: initial dynamic frequency sweep (DFS *before*), series of start-up-shear and stress relaxation after flow cessation at increasing shear rates as indicated in the legend, and final DFS (DFS *after*).

Typical shear start-up curves, shear stress growth function, $\sigma^+$, versus strain, $\gamma$, are displayed in Fig. 2 for sample p4. After a monotonic increase, the stress attains a maximum value, $\sigma_{max}$, before reaching a steady state, $\sigma_{steady}$, with eventually an overshoot characterized by $\sigma_{max} >$ $\sigma_{steady}$ (Fig. 2a). At long times, i.e. for the lowest shear rate (0.1 s$^{-1}$), ageing is evidenced by a weak increase of the stress, while for the highest shear rates, the measured stress may be affected by instabilities (edge fracture and/or shear banding), preventing a true steady state stress to be attained. The shear rates, for which a true steady state is not reached, were excluded from analysis. When the shear stress growth coefficient, $\eta^+ = \frac{\sigma^+}{\dot{\gamma}}$, is plotted as a function of time, $t$, (time 0 corresponds to the time at which the shear rate is applied), data at short times acquired with different prescribed shear rates all superimpose on the linear envelope (see Fig. SI3a). Interestingly, we note that the maximum shear stress value is close to the value measured at the end of the linear regime (see Fig. SI3b). Accordingly, the evolution of $\sigma_{max}$ as a function of the shear rate $\dot{\gamma}$ can be plotted on a single master curve gathering the data for all samples and using $\tau_z$ and $b_L$ as shifting factors (Fig. 2b). Hence, the maximum shear stress reached at a given shear rate for gelling systems with different extents of gelation can be estimated using linear viscoelasticity. Figure 2c displays the increase of $\sigma_{steady}$ with the shear rate for the different samples. At low shear rates, a power law evolution is measured. We find that the power law exponent, κ, decreases with the extent of gelation, from κ=1 for samples p1 and p2, indicating a Newtonian fluid behavior, down to κ=0.5 for sample p6 (inset Fig. 2c). κ<1 indicates shear thinning materials, which can be rationalized by the Soft Glassy Rheology model (SGR)[25]. The SGR model indeed describes materials with mesoscopic disorder, which undergo slow ageing. It considers mesoscopic elements whose dynamics consists of independent hopping between local traps. Hopping is controlled by an effective temperature parameter x=T$_{eff}$/T$_g$, that represents the system noise that is responsible for the spontaneous relaxation dynamics. Here, T$_{eff}$ is an effective temperature and T$_g$ is the glass transition temperature. The SGR model combines the hopping dynamics with the buildup of local elastic shear strains in the mesoscopic elements and predicts for power law fluids, 1<x<2, $\sigma(\dot{\gamma} \to 0) \propto \dot{\gamma}^{x-1}$. Our measurements which show a decrease of the exponent κ with the slowing down of the sample dynamics (increase of $\tau_z$) is in qualitative accordance with the SGR models that predicts a decrease of the exponent (x-1) as the samples approaches the glass transition (x=1).

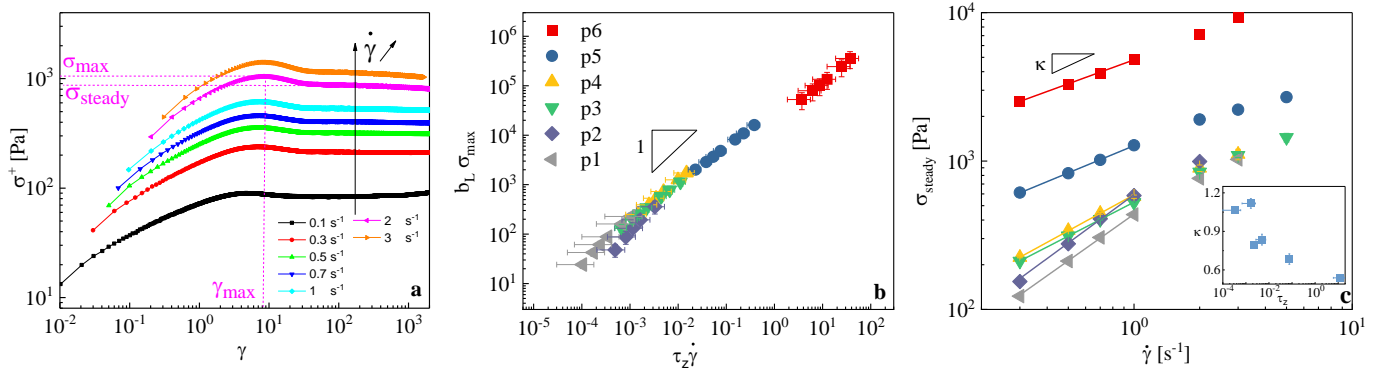

**Figure 2.** (a) Shear stress growth function versus strain, at different shear rates, as indicated in the legend, for sample p4. $\sigma_{max}$, $\sigma_{steady}$ and $\gamma_{max}$ are indicated for $\dot{\gamma}=2s^{-1}$. (b) Master curve for the maximum stress, $b_L \sigma_{max}$ vs $\tau_z \dot{\gamma}$, with $b_L$ and $\tau_z$ the average shifting factors as determined from the linear viscoelasticity measurements (Fig. 1). (c) $\sigma_{steady}$ vs $\dot{\gamma}$ for all samples. Full line: best fit of experimental data in the low shear rate regime with a power law function. Inset: power law exponent, κ, as a function of $\tau_z$.

To better understand the non-linear behavior of the gelling systems, we focus on the amplitude of the overshoot $\frac{\sigma_{max}}{\sigma_{steady}}$, which is commonly used to characterize the transient rheology of polymer systems[26,27]. We find that the amplitude increases systematically with the shear rate and the extent of gelation (Fig. 3a). Interestingly, a unique master curve can be built by normalizing the shear rate by a characteristic shear rate, $1/\alpha$, evidencing two regimes (Fig. 3b): at low shear rates, $\frac{\sigma_{max}}{\sigma_{steady}}$ is constant and close to one, while at high shear rates $\frac{\sigma_{max}}{\sigma_{steady}}$ increases as a power law of the shear rate with an exponent 0.25. Note that similar two-regime master curves were previously obtained for polymer melts[27], metal-ligand networks[28] and soft particle glasses[29], and the time associated with the change of regime was the disentanglement time, the bond lifetime, and the cage rearrangement time, respectively. Hence, the physical origin of the stress overshoot is generally associated to an elastic stress stored in the sample before structural rearrangements occur. Here, the transition defines a sample characteristic time for each pre-gel, $\alpha$, which is different from the time defined in the linear regime, $\tau_z$. However, remarkably, these times are related together by a power-law $\alpha \propto \tau_z^n$ with $n = 0.4$, which is the critical exponent defined in the linear regime (Fig. 3c). Therefore, the characteristic time of the stress overshoot is a function of the two parameters defining the whole distribution of relaxation times in the linear regime ($n$ and $\tau_z$). According to the near-critical gel theory[8], $\tau_z^n$ is proportional to $\frac{\tau_z}{\eta_0}$, a quantity which is inversely proportional to the energy stored by a pre-gel in a creep experiment. Remarkably, the inverse relationship between $\alpha$ and the elastic energy stored by the system recalls results obtained with soft particle glasses[29,30]. In these works, the authors indeed found that the local relaxation time of soft particles, which is pertinent for scaling several rheological data such as flow curves[30] and stress overshoot in shear start-up experiments[29], is proportional to $\eta_s/G_e$ with $\eta_s$ the solvent viscosity and $G_e$ the sample elastic modulus.

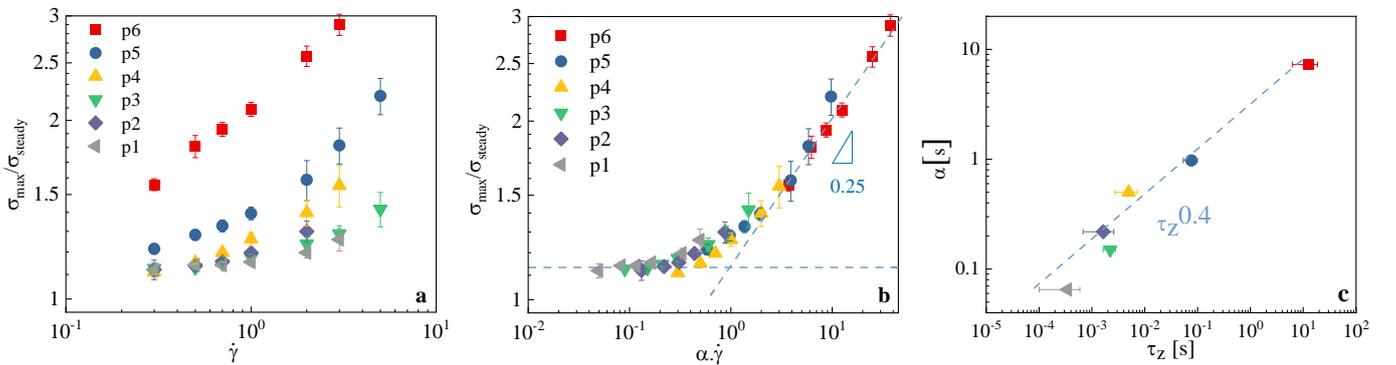

**Figure 3.** (a) Stress overshoot, $\frac{\sigma_{max}}{\sigma_{steady}}$, as a function of the shear rate for the different samples as indicated in the legend. (b) Master curve of the stress overshoot as obtained using α as a horizontal

shifting factor, for the same data as in (a). (c) Evolution of α with $\tau_z$. Symbols are experimental data points, and the dashed line is a power law fit yielding an exponent 0.4.

The analogy between pre-gels and glasses stems from the slowing down of the characteristic dynamics of both classes of systems. To deepen this analogy, we investigate the stress relaxation upon flow cessation. We measure that, for all shear rates and all samples, the shear stress decay function measured after flow cessation relaxes to 0, as expected for liquid-like gelling samples before the gel point (Fig. 4a). Moreover, the stress relaxation curves measured after different shear rates display similar shapes as evidenced by the master curve built for a given sample using $\dot{\gamma}t$ as *x*-axis (Fig. 4b), indicating that the characteristic relaxation time is inversely proportional to the shear rate. Here *t*=0 corresponds to the time when the shear rate is set to 0. This scaling and the power law decay of the stress with time have been previously observed for hard[31] and soft glasses[32] close to the glass transition, and envisioned by the SGR model[25]. SGR indeed predicts $\sigma^-(t) \sim (\dot{\gamma}t)^{-x}$ for $1 < x < 2$ [25]. In our experiments, the power law exponent -δ is close to -1, and weakly increases from -1.2 to -0.6 as the extent of gelation p increases (inset of fig.4b), indicating a slower stress relaxation as the sample approaches percolation, in qualitative accordance with the SGR model. In addition, according to this model, the $\dot{\gamma}t$ scaling would reflect the glassy nature of the sample (mesoscopic dynamics) by contrast to a *t*-dependence expected for a dynamic dominated by microscopic time scales. Furthermore, the shear rate dependence of the relaxation was attributed to a slow decay of mesoscopic structural anisotropies induced by the flow. To investigate the role of the extent of gelation on the relaxation dynamics, we define a characteristic relaxation time, $\tau_R$, as the time when 90% of the initial stress has been relaxed. This time decreases with the shear rate and increases with p (Fig. 4c). Interestingly, using an analysis inspired from works on polymer melts[27] and soft particle glasses[33], we find that data from the different samples collapse on a unique master curve when the dimensionless relaxation time, $\tau_R/\tau_z$, is plotted as a function of the Weissenberg number, $\dot{\gamma}\tau_z$ (Fig. 4d). Hence, the mesoscopic dynamics responsible for the stress relaxation after flow cessation scales with $\tau_z$, the sample longest relaxation time that probes the extent of gelation in the linear regime.

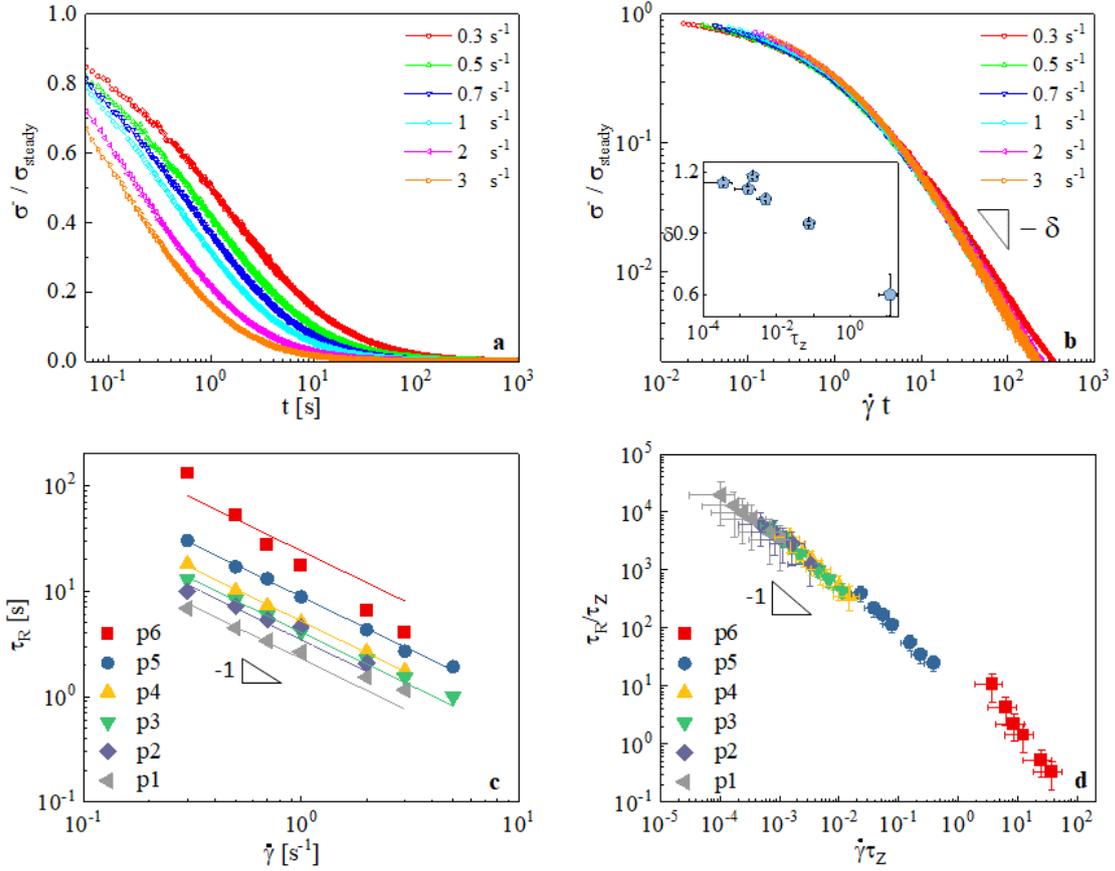

**Figure 4.** Stress relaxation upon cessation of steady flow. (a) Normalised shear stress decay function for sample p4 after shear start-up at different shear rates $\dot{\gamma}$ as indicated in the legend. (b) Same data as in (a) but plotted in log-log scale as a function of $\dot{\gamma}$t. The inset displays the power law exponent δ of the long-time decay as a function of $\tau_z$. (c) Characteristic relaxation time $\tau_R$ (see main text) as a function of $\dot{\gamma}$ for samples with different extents of gelation. Symbols are experimental data points, and the continuous lines are power-law fits imposing an exponent of -1 (d) Evolution of the dimensionless time, $\frac{\tau_R}{\tau_z}$ with the Weissenberg number, $\dot{\gamma}\tau_z$.

To summarize, we have studied model gluten samples that can reach stable states at different extents of gelation allowing long experiments to be conducted to investigate the response of near-critical gelling systems beyond the linear regime. We have shown that the samples display self-similarities when submitted to non-linear shear start-up deformations. For samples close to the gel point and at high shear rates, a stress overshoot is measured before reaching a steady flow. The stress overshoot amplitude of the different samples can be plotted on a single master curve using shifting factors inversely proportional to the linear viscoelastic moduli of the samples, as previously evidenced for colloidal jammed suspensions[29]. In addition, stress relaxations after flow cessation display power law decays with a shear rate dependence that is characteristic of materials driven by a structural disorder at the mesoscopic scale[25]. The gelation-induced slowing down of the sample dynamics leads to rheological features in the non-linear regime that are comparable to those of systems approaching a glass or jamming transition. The comparison of dynamical and rheological consequences of gelation with that of glass transition was previously proposed by Segrè et al[34] and Winter[35]. Our work illustrates the analogy between gelation and glass transition in non-linear rheology.

**Experimental methods:**

Freeze-dried gluten extracts with different glutenin contents were obtained according to a protocol described elsewhere[36]. Samples were prepared at 500g/L in milliQ water with 0.1% sodium azide. For some samples, N-ethyl maleimide (NEM) was previously added in the solvent. The samples' composition is given in Table 1. Dispersions of gluten powder in solvent were manually stirred for 3 minutes, and then stored for 5 days at room temperature, prior to measurements. Rheology was measured using a MCR 302 rheometer (Anton Paar, Austria), operated in strain-controlled mode. A rough (~6 μm in height), cone-plate geometry (8 mm-diameter cone, 3 degrees) was used to minimize sample slip. Samples together with the geometry were immersed in a bath of silicon oil to avoid solvent evaporation. After sample loading, several tens of minutes of rest are given to the samples to reach a negligible normal force and relax the residual stress induced by loading. The linear viscoelasticity (LVE) was measured through dynamic frequency sweeps from 100 rad/s to 0.01 rad/s, using a strain $\gamma_0$=0.1, which is within the linear regime identified by means of a dynamic strain sweep. The nonlinear viscoelasticity was measured through a shear start-up protocol, which consists in applying a fixed steady shear rate, $\dot{\gamma}$, for a certain time, $t$, and monitoring the evolution of the shear stress growth function $\sigma^+(t,\dot{\gamma})$. We imposed different shear rates, $\dot{\gamma}$, from $0.1\ s^{-1}$ to $10\ s^{-1}$ for a fixed total strain of 2000, followed by a relaxation period ($\dot{\gamma}$ is set to 0) of 2400 s between each shear start-up run. The series were bordered by LVE tests (Fig 1c).

**Supporting information:**

Tan delta, Cole-Cole plots, Shear stress growth coefficient and complex viscosity, correlation between the onset of the deviation from the linear regime and the maximum stress.

**Acknowledgments:**

This work was financially supported by the French National Agency that funded the young researcher project entitled Elastobio, grant: ANR-18-CE06-0012-01. We acknowledge Pr. V. Trappe from the University of Fribourg (CH) for fruitful discussions.

# Supporting Information

In order to better appreciate the near critical gel regime of samples, linear viscoelasticity data are represented as tan delta vs frequency in Fig. SI 1, and as Cole-Cole plots in Fig. SI 2. Raw data measured before and after the shear start-up protocol are plotted in figures a), and shifted data are given in figures b). We use here the numerical values for $\tau_Z$ and $b_L$ as those used in Fig.1 of the main manuscript.

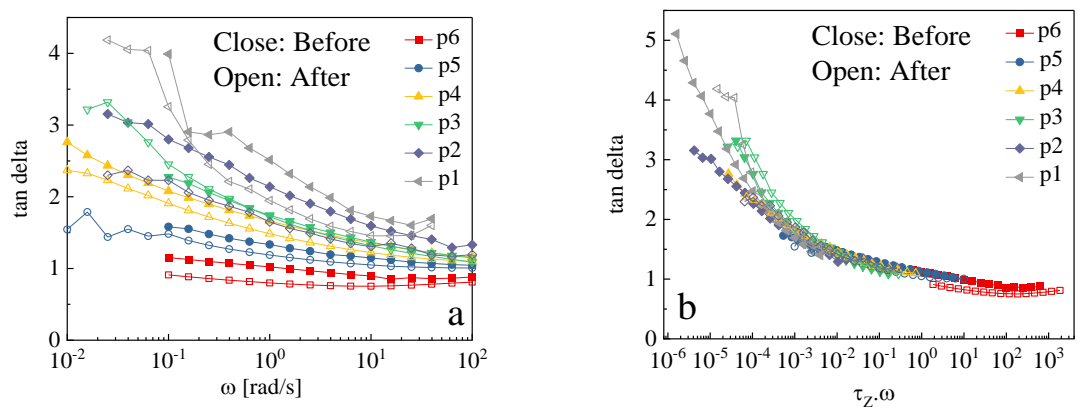

**Figure SI 1.** Tan delta=G"/G' representation of linear viscoelasticity data measured before and after the shear start-up protocol. (a) Raw data; (b) Master curve obtained using $\tau_z$ as shifting factor.

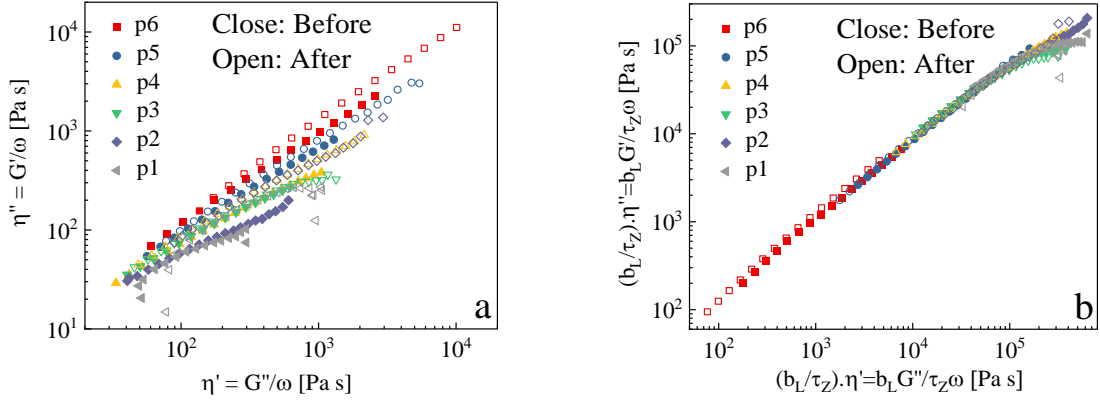

**Figure SI 2.** Cole-Cole representation of linear viscoelasticity data measured before and after the shear start-up protocol. (a) Raw data; (b) Master curve obtained using $\tau_z$ and $b_L$ as shifting factors. In (a) $\eta'' = \frac{G''}{\omega}$ is plotted as a function of $\eta' = \frac{G'}{\omega}$, and in (b) the complex viscosity is normalized by $\frac{\tau_z}{b_L}$.

To appreciate the extent of the linear regime in start-up shear data, the shear stress growth coefficient $\eta^+ = \frac{\sigma}{\dot{\gamma}}$ is plotted versus time (time 0 corresponds to the time at which the shear rate is applied) and superimposed with linear data. Linear data are computed through a direct transformation of the dynamic linear data by applying the Cox-Merz rule $\eta(\dot{\gamma}) = \eta^*(\omega)|_{\omega=\dot{\gamma}}$[25] in conjunction with the Gleissle relationship $\eta^+(t) = \eta(\dot{\gamma})|_{\dot{\gamma}=1/t}$[26], where the complex viscosity reads $\eta^* = \frac{\sqrt{G'^2 + G''^2}}{\omega}$. At short time, shear startup data are comparable to the linear envelope. The onset of the deviation from the linear regime is measured at a stress value close to the maximum value as evidenced in fig. SI 3b.

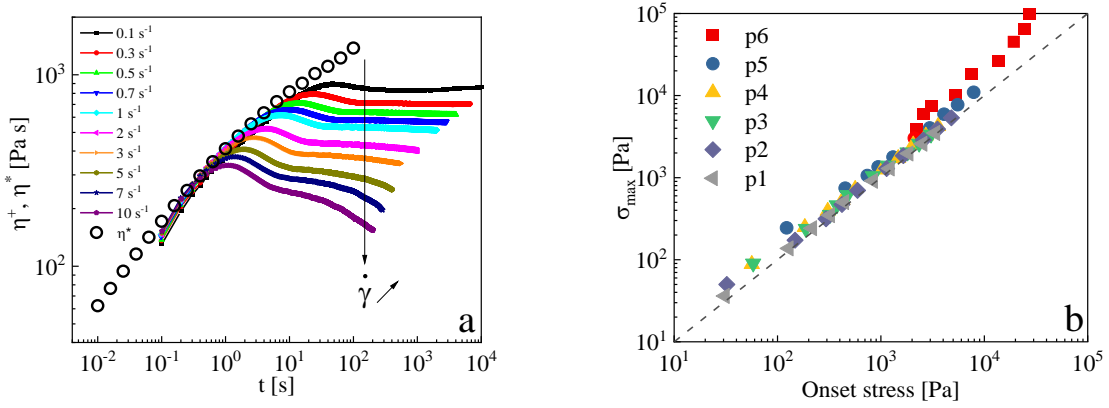

**Figure SI 3.** (a) Shear stress growth coefficient $\eta^+ = \frac{\sigma}{\dot{\gamma}}$ versus time (colored full symbols), for sample p4 for the same shear rates as in Figure 2(a), superimposed with linear data, complex viscosity $\eta^*$ versus $1/\omega$ (empty circles). (b) Correlation between the maximum stress and the stress associated with the onset on the deviation from the linear envelope for all samples and all shear rates. Colored symbols correspond to experimental data points for different samples, as indicated in the legend, and the dashed line indicates the equality between the two stresses.